\documentclass[reprint,superscriptaddress,aps]{revtex4-1}
\usepackage{amsmath}
\usepackage{amssymb}
\usepackage{bm}
\usepackage{braket}
\usepackage{color}
\usepackage[varg]{txfonts}
\usepackage{varwidth}
\usepackage{dcolumn}
\usepackage[breaklinks,colorlinks=true,linkcolor=blue,urlcolor=cyan,citecolor=blue]{hyperref}
\usepackage{graphicx}
\usepackage{here}

\begin{document}

\title{Complex magnetic phase diagram with a small phase pocket in a three-dimensional frustrated magnet CuInCr$_{4}$S$_{8}$}

\author{M. Gen}
\email{masaki.gen@riken.jp}
\affiliation{Institute for Solid State Physics, The University of Tokyo, Kashiwa 277-8581, Japan}
\affiliation{Department of Advanced Materials Science, The University of Tokyo, Kashiwa 277-8561, Japan}

\author{H. Ishikawa}
\affiliation{Institute for Solid State Physics, The University of Tokyo, Kashiwa 277-8581, Japan}

\author{A. Ikeda}
\affiliation{Institute for Solid State Physics, The University of Tokyo, Kashiwa 277-8581, Japan}
 
\author{A. Miyake}
\affiliation{Institute for Solid State Physics, The University of Tokyo, Kashiwa 277-8581, Japan}

\author{Z. Yang}
\affiliation{Institute for Solid State Physics, The University of Tokyo, Kashiwa 277-8581, Japan}

\author{Y. Okamoto}
\affiliation{Department of Applied Physics, Nagoya University, Nagoya 464-8603, Japan}

\author{M. Mori}
\affiliation{Department of Applied Physics, Nagoya University, Nagoya 464-8603, Japan}

\author{K. Takenaka}
\affiliation{Department of Applied Physics, Nagoya University, Nagoya 464-8603, Japan}

\author{H. Sagayama}
\affiliation{Institute of Materials Structure Science, High Energy Accelerator Research Organization, Tsukuba 305-0801, Japan}

\author{T. Kurumaji}
\affiliation{Department of Advanced Materials Science, The University of Tokyo, Kashiwa 277-8561, Japan}

\author{Y. Tokunaga}
\affiliation{Department of Advanced Materials Science, The University of Tokyo, Kashiwa 277-8561, Japan}

\author{T. Arima}
\affiliation{Department of Advanced Materials Science, The University of Tokyo, Kashiwa 277-8561, Japan}

\author{M. Tokunaga}
\affiliation{Institute for Solid State Physics, The University of Tokyo, Kashiwa 277-8581, Japan}

\author{K. Kindo}
\affiliation{Institute for Solid State Physics, The University of Tokyo, Kashiwa 277-8581, Japan}

\author{Y. H. Matsuda}
\affiliation{Institute for Solid State Physics, The University of Tokyo, Kashiwa 277-8581, Japan}

\author{Y. Kohama}
\affiliation{Institute for Solid State Physics, The University of Tokyo, Kashiwa 277-8581, Japan}

\begin{abstract}

Frustrated magnets with a strong spin-lattice coupling can show rich magnetic phases and the associated fascinating phenomena.
A promising platform is the breathing pyrochlore magnet CuInCr$_{4}$S$_{8}$ with localized $S=3/2$ Cr$^{3+}$ ions, which is proposed to be effectively viewed as an $S=6$ Heisenberg antiferromagnet on the face-centered-cubic lattice.
Here, we unveil that CuInCr$_{4}$S$_{8}$ exhibits a complex magnetic phase diagram with a small phase pocket ({\it A} phase) by means of magnetization, magnetostriction, magnetocapacitance, and magnetocaloric-effect measurements in pulsed high magnetic fields of up to 60~T.
Remarkably, the appearance of {\it A} phase is accompanied by anomalous magnetostrictive and magnetocapacitive responses, suggesting the emergence of helimagnetism in contrast to the neighboring commensurate magnetic phases.
Besides, the high-entropy nature is confirmed in the high-temperature side of {\it A} phase.
These features are potentially related to a thermal fluctuation-driven multiple-$q$ state caused by the magnetic frustration, which has been theoretically predicted but yet experimentally undiscovered in insulating magnets.

\end{abstract}

\date{\today}
\maketitle

\section{\label{Sec1}Introduction}

Magnetic frustration is a key ingredient for the emergence of exotic magnetic states in condensed matters.
One promising route to induce strong frustration is to arrange localized spins on a crystal lattice involving triangular motifs, as represented by the Heisenberg antiferromagnet on the pyrochlore lattice, a three-dimensional network of corner-sharing tetrahedra.
In the classical-spin picture, competing nearest-neighbor (NN) exchange interactions arising from geometrical constraints prevent the development of a magnetic long-range order even at low temperatures \cite{1998_Moe}.
In real systems, however, the lattice degrees of freedom often come into play to lift the macroscopic degeneracy \cite{2002_Tch, 2004_Pen}, yielding a rich magnetic phase diagram with spin-lattice-coupled ordered states in the parameter space of temperature, magnetic field, and pressure.

Chromium spinels {\it A}Cr$_{2}${\it X}$_{4}$ ({\it A}=Mg, Zn, Cd, Hg; {\it X}=O, S, Se) are known as typical insulating frustrated magnets to realize versatile magnetic phenomena caused by the interplay between the magnetic frustration and the spin-lattice coupling (SLC): e.g., a 1/2-magnetization plateau \cite{2006_Ued, 2008_Koj, 2011_Miy, 2014_Miy}, negative thermal expansion \cite{2007_Hem, 2013_Kit, 2019_Ros}, and spin-driven electric polarization \cite{2006_Web, 2008_Mur, 2020_Ros}.
Here, tetrahedrally coordinated nonmagnetic {\it A}$^{2+}$ ions and octahedrally coordinated $S=3/2$ Cr$^{3+}$ ions form a diamond and pyrochlore network, respectively.
The NN Cr--Cr exchange interaction arises from the antiferromagnetic (AFM) direct exchange due to the overlapping of {\it t}$_{2g}$ orbitals and the ferromagnetic (FM) superexchange via Cr--{\it X}--Cr path.
In oxides, the former contribution is predominant because of the short Cr--Cr length \cite{2008_Yar}, so that the system can be approximated to the NN Heisenberg antiferromagnet with geometrical frustration.
In sulfides and selenides, by contrast, the superexchange contribution exceeds the direct one, and further-neighbor exchange interactions via Cr--{\it X}--{\it X}--Cr and Cr--{\it X}--{\it A}--{\it X}--Cr paths also become strong \cite{2008_Yar}, so that the magnetic properties are governed by bond frustration \cite{2011_Tsu}.

\begin{figure}[t]
\centering
\includegraphics[width=\linewidth]{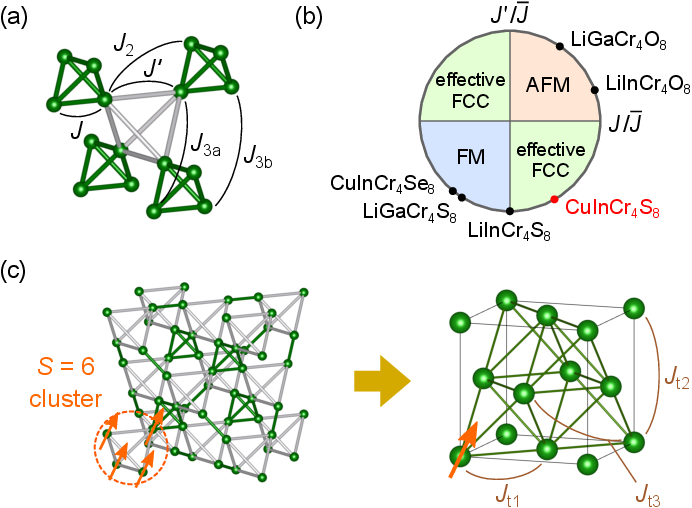}
\caption{(a) Breathing pyrochlore lattice comprising magnetic ions with two kinds of the NN exchange interactions, $J$ and $J'$, in the small and large tetrahedra, respectively, and further-neighbor exchange interactions up to the third NN. (b) Circular representation of the magnetic models of several breathing pyrochlore Cr spinels as a function of $J$ and $J'$ normalized to ${\overline J}\equiv \sqrt{J^{2}+J'^{2}}$, which are obtained from the previous DFT calculation \cite{2019_Gho}. (c) Effective magnetic model of CuInCr$_{4}$S$_{8}$ mapped on the face-centered-cubic lattice with the NN AFM exchange interaction $J_{\rm t1} \equiv J+4J_{2}+2(J_{\rm 3a}+J_{\rm 3b})$ and the second and third NN interactions, $J_{\rm t2}$ and $J_{\rm t3}$, respectively. The illustrations in (a) and (c) are drawn with VESTA software \cite{2011_Mom}.}
\label{Fig1}
\end{figure}

Recently, {\it A}-site ordered Cr spinels {\it A}{\it A'}Cr$_{4}${\it X}$_{8}$ ({\it A}=Li, Cu; {\it A'}=Ga, In), where nonmagnetic monovalent {\it A}$^{+}$ and trivalent {\it A'}$^{3+}$ cations are regularly arranged, have attracted new attention in terms of the ground state control \cite{2013_Oka, 2014_Tan, 2015_Nil, 2016_Sah, 2019_Gen, 2020_Kan, 2018_Oka, 2020_Gen, 2021_Gao, 2020_Pok, 2019_Gho}.
The difference in their chemical pressure renders the Cr$_{4}$ tetrahetra expanding and contracting alternately, resulting in the so-called breathing pyrochlore lattice with nonequivalent NN exchange interactions $J$ and $J'$ in the small and large tetrahedra, respectively [Fig.~\ref{Fig1}(a)].
A large majority of previous studies have focused on Li(Ga, In)Cr$_{4}$O$_{8}$ with $J>J'>0$ (AFM) \cite{2013_Oka, 2014_Tan, 2015_Nil, 2016_Sah, 2019_Gen, 2020_Kan}, where the system can be viewed as an intermediate state between the regular pyrochlore lattice ($J'/J=1$) and the isolated tetrahedral clusters ($J'/J=0$).
For small values of $J'/J$, the spin-gapped behavior associated with the tetramer singlet formation rather than the classical spin-liquid behavior can emerge, as observed in LiInCr$_{4}$O$_{8}$ \cite{2013_Oka, 2014_Tan, 2015_Nil}.
Furthermore, it has been theoretically suggested that a variety of magnetic long-range orders can appear depending on the value of $J'/J$ and the strength of the SLC \cite{2019_Aoy, 2021_Aoy}, which seems relevant to the magnetic structures on Li(Ga, In)Cr$_{4}$O$_{8}$ at low temperatures \cite{2015_Nil, 2016_Sah}.

In the breathing pyrochlore Cr spinels, not only the magnitude but the sign of $J$ and $J'$ can be flexibly tuned by the chemical substitution, which can bring about a dramatic modification in the magnetic model rather than a perturbative change.
Of particular interest is CuInCr$_{4}$S$_{8}$ \cite{2018_Oka, 2020_Gen, 2021_Gao}, where the special condition of $J>0$ (AFM) and $J'<0$ (FM) is satisfied due to the delicate balance between the direct and superexchange interactions [Fig.~\ref{Fig1}(b)]; the exchange parameters at room temperature are estimated to $J=14.7$~K, $J'=-26.0$~K, $J_{2}=1.1$~K, $J_{\rm 3a}=6.4$~K, and $J_{\rm 3b}=4.5$~K by the density-functional-theory calculation [Fig.~\ref{Fig1}(a)] \cite{2019_Gho}.
Since the FM $J'$ does not compete with any other exchange interactions listed above, magnetic moments on each large Cr$_{4}$ tetrahedron are aligned ferromagnetically at low temperatures, as evidenced by the observation of the cluster excitation using inelastic neutron scattering \cite{2021_Gao}.
The magnetism of CuInCr$_{4}$S$_{8}$ can hence be mapped onto $S=6$ spins on the face-centered-cubic (FCC) lattice with the NN AFM interaction $J_{\rm t1} \equiv J+4J_{2}+2(J_{\rm 3a}+J_{\rm 3b})$ [Fig.~\ref{Fig1}(c)].
Since the geometrical frustration remains in this effective FCC lattice, a robust 1/2-magnetization plateau with a 3-up-1-down magnetic structure ({\it C} phase) appears via the exchange-striction mechanism \cite{2020_Gen}, as in Cr spinel oxides \cite{2006_Ued, 2008_Koj, 2011_Miy, 2014_Miy, 2019_Gen}.
On the other hand, the presence of an intermediate-field phase ({\it Y} phase) in a wide field range between the low-field AFM phase ({\it X} phase) and {\it C} phase is a distinctive feature of this compound, which seems to owe to the peculiar combination of the emergent FCC lattice and the SLC \cite{2020_Gen}.
The previous focus on the in-field properties of CuInCr$_{4}$S$_{8}$ was limited to the ground state \cite{2018_Oka, 2020_Gen}, while the effect of thermal fluctuations on the stability of {\it C} and {\it Y} phases have remained unexplored.

Here, we report on a fascinating magnetic-field-vs-temperature ($H$-$T$) phase diagram of CuInCr$_{4}$S$_{8}$, revealed by means of magnetization, magnetostriction, magnetocapacitance, and magnetocaloric-effect (MCE) measurements in pulsed high magnetic fields of up to 60~T.
The most striking finding is the existence of a small phase-pocket ({\it A} phase), which extends from the phase boundary with the paramagnetic phase towards lower temperatures instead of {\it Y} phase.
Remarkably, the transition from {\it X} to {\it A} phase is accompanied by anomalous magnetostrictive and magnetocapacitive responses.
These observations suggest the appearance of helimagnetism in {\it A} phase through a commensurate-to-incommensurate phase transition.
Moreover, the high-entropy nature is confirmed in the high-temperature side of {\it A} phase.
Given the similarity of the identified $H$-$T$ phase diagram to that of a number of skyrmion hosts \cite{2009_Muh, 2012_Sek, 2021_Tok}, the emergence of a multiple-$q$ vortex-like magnetic structure is also expected.
We propose that the unique magnetic properties of CuInCr$_{4}$S$_{8}$ could be ascribed to the cooperative effect of the strong SLC and the breathing bond-alternation introducing the FCC-lattice magnetism.

\section{\label{Sec2}Experimental methods}

The magnetic ordering temperatures of CuInCr$_{4}$S$_{8}$ reported so far range from 26 to 35~K \cite{2021_Gao, 1970_Pin, 1980_Plu, 2016_Ami}.
Our previous polycrystalline samples (Sample no.~2) exhibit a magnetic phase transition at $T_{\rm N} \approx 32$~K \cite{2018_Oka}, which is slightly different from any values reported by other groups.
We prepared new polycrystalline batches (Sample no.~1) with the solid state reaction as in Ref.~\cite{2018_Oka}, and found a higher transition temperature as $T_{\rm N} \approx 35$~K [Figs.~\ref{Fig2}(a) and \ref{Fig2}(b)], in accordance with that reported in Refs.~\cite{2021_Gao, 1980_Plu}.

To compare the sample quality between Samples no.~1 and no.~2, powder x-ray diffraction (XRD) patterns were obtained using synchrotron x-ray at BL-8A at Photon Factory.
The Rietveld analyses confirm the cubic $F{\overline 4}3m$ structure with negligible crystallographic disorder and/or defects in the nonmagnetic Cu$^{+}$ and In$^{3+}$ sites and the ligand S$^{2-}$ sites for both samples, while some differences in the crystallographic parameters are found between them (Appendix~\ref{SecA}).
Furthermore, a series of high-field experiments revealed that field-induced phase transitions are much sharper in Sample no.~1 than in Sample no.~2, suggesting that Sample no.~1 is of higher quality, i.e., the Cr-based pyrochlore network may be better breathing.\
The results for Sample no.~1 are shown in the main text, while those for Sample no.~2 are shown in Appendixes~\ref{SecB} and \ref{SecC}.

Magnetization up to 7~T was measured using a commercial magnetometer (MPMS, Quantum Design).
Magnetization up to $\sim$60~T was measured by the conventional induction method using a coaxial pickup coil in a non-destructive (ND) pulsed magnet ($\sim$4~ms duration) at the Institute for Solid State Physics (ISSP), University of Tokyo, Japan.
Longitudinal magnetostriction up to $\sim$60~T was measured with the optical fiber-Bragg-grating (FBG) technique in a ND pulsed magnet ($\sim$36 ms duration) at ISSP \cite{2018_Ike}.
The fiber was adhered to a rod-shaped sintered sample with epoxy Stycast1266.
The distortion $\Delta L/L$ was detected by the optical filter method with a resolution of $\sim$10$^{-6}$.
The thermal expansion at zero field was measured using the same setup during a natural temperature rise from 2 to 60~K.
The absolute value of $\Delta L/L$ is not calibrated, so that the displayed value of $\Delta L/L$ would be smaller than actual lattice constant change.
Dielectric constant at zero field was measured at a frequency of 10~kHz by using an LCR meter (E4980A, Agilent) in a commercial cryostat equipped with a superconducting magnet (PPMS, Quantum Design).
Magnetocapacitance along the field direction ($E \parallel B$) up to $\sim$60~T was measured at a frequency of 50~kHz by using a capacitance bridge (1615-A, General Radio) in a ND pulsed magnet ($\sim$36~ms duration) at ISSP \cite{2020_Miy}.
Silver paste was painted on the top and bottom surfaces to form electrodes for a disk-shaped sintered sample.
Heat capacity at zero field was measured by the thermal relaxation method using a PPMS.
The magnetocaloric effect up to $\sim$60~T was measured under both the nonadiabatic and adiabatic conditions in a ND pulsed magnet ($\sim$36~ms duration) at ISSP.
A sensitive Au$_{16}$Ge$_{84}$ film thermometer was sputtered on the surface of a disk-shaped sintered sample, which was calibrated by a commercial Cernox 1030 thermometer \cite{2013_Kih}.

\section{\label{Sec3}Results and discussions}

\subsection{\label{Sec3_1} Magnetic properties of the low-field phase ({\it X} phase)}

\begin{figure}[t]
\centering
\includegraphics[width=0.75\linewidth]{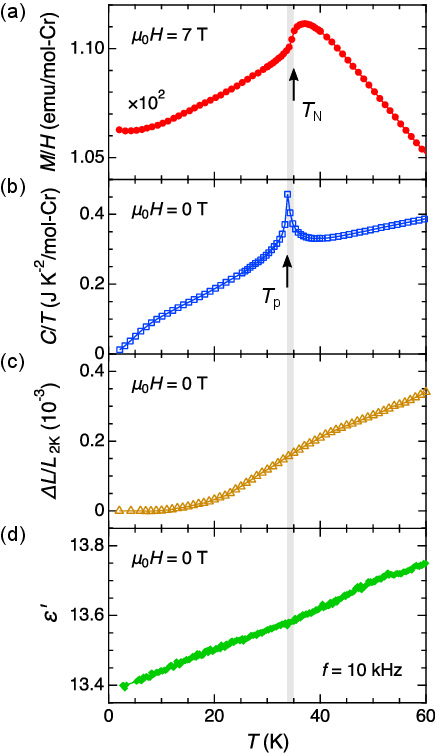}
\caption{(a) Magnetization measured at 7~T after zero-field cooling. (b) Specific heat divided by temperature at 0~T. (c) Thermal expansion compared to the 2-K data at 0~T. (d) Dielectric constant at a frequency of 10~kHz at 0~T. All the measurements were performed in a warming process.}
\label{Fig2}
\end{figure}

We first mention notable aspects of the low-field phase ({\it X} phase) of CuInCr$_{4}$S$_{8}$.
Cr spinels typically undergo a first-order phase transition into an AFM 2-up-2-down collinear structure \cite{2015_Nil, 2016_Sah, 2007_Mat, 2009_Ji} or an incommensurate helical structure \cite{2008_Mur, 2011_Tsu} accompanied by the crystal symmetry lowering due to the SLC.
For CuInCr$_{4}$S$_{8}$, the appearance of {\it X} phase is associated with a drop in the magnetic susceptibility $M/H$ at $T_{\rm N}=35$~K [Fig.~\ref{Fig2}(a)] and a peak in the specific heat $C$ at $T_{\rm p}=34$~K [Fig.~\ref{Fig2}(b)].
A recent neutron diffraction study has revealed that the crystal structure remains the cubic $F{\overline 4}3m$ symmetry down to 2~K \cite{2021_Gao}.
Our further experiments also support the uniqueness of {\it X} phase; neither the normalized thermal expansion $\Delta L/L_{\rm 2K}$ nor the dielectric constant $\varepsilon'$ exhibit any sharp anomalies at $T_{\rm N}$ [Figs.~\ref{Fig2}(c) and \ref{Fig2}(d)], unlike in other Cr spinels \cite{2007_Hem, 2013_Kit, 2019_Ros, 2016_Sah, 2020_Kan}.
Furthermore, no hysteretic behavior is observed in the $M/H$-$T$ curve (not shown), indicating that the phase transition at $T_{\rm N}$ is of the second order, and the SLC does not seem to be responsible for the appearance of {\it X} phase.

In order to understand the characteristic of {\it X} phase, it is useful to consider the $J_{\rm t1}$--$J_{\rm t2}$--$J_{\rm t3}$ model on the effective FCC lattice, where $J_{\rm t2}$ and $J_{\rm t3}$ stand for the second and third NN interactions in the effective FCC model [Fig.~\ref{Fig1}(a)].
Theoretically, two kinds of commensurate AFM order, Type-I with {\bf q}=(1 0 0) and Type-III with {\bf q}=(1 1/2 0), appear at zero field for $J_{\rm t2}<4J_{\rm t3}$ and $J_{\rm t2}>4J_{\rm t3}$ respectively, if $|J_{\rm t2}|, |J_{\rm t3}| \ll J_{\rm t1}$ \cite{2020_Bal}., respectively
Note that $J_{\rm t2}$ and $J_{\rm t3}$ arise from the fourth or higher order NN exchange interactions in the original breathing pyrochlore lattice, and should be much weaker than $J_{\rm t1}$ in CuInCr$_{4}$S$_{8}$.
Indeed, Type-I order has been confirmed by powder neutron diffraction, and the observed magnon dispersion is accountable by incorporating FM $J_{\rm t2}$ ($<0$) \cite{2021_Gao, 1971_Plu}.
Although the local spin configuration of {\it X} phase has not been fully identified, an ``all-in-all-out''-like one \cite{1971_Plu} and a coplanar one in which the two AFM spin pairs are canted to each other \cite{2021_Gao} have been proposed as candidates rather than the 2-up-2-down collinear one.
This is curious because the SLC and thermal fluctuations tend to favor a collinear spin configuration out of the continuous degeneracy regarding the directions of four inequivalent spins in Type-I order \cite{1987_Hen, 2015_Ben}.
Moreover, neither of the proposed magnetic structures can be explained just by considering the Dzyaloshinskii-Moriya (DM) interactions which should exist to a greater or lesser extent in the breathing pyrochlore lattice \cite{1957_Dzy, 1960_Mor, 2005_Elh} (Appendix~\ref{SecD}).
We note that a small singularity at $T_{\rm p}$ and a hump structure around 10~K in the $C/T$-$T$ curve [Fig.~\ref{Fig2}(b)] suggest the existence of a large magnetic entropy even below $T_{\rm N}$ (Appendix~\ref{SecE}).
These features are unique to CuInCr$_{4}$S$_{8}$ among Cr spinels \cite{2013_Kit, 2013_Oka, 2018_Oka, 2013_Kem} and would stem from the magnetic frustration inherent in the effective FCC-lattice Heisenberg antiferromagnet.

\begin{figure*}[t]
\centering
\includegraphics[width=0.88\linewidth]{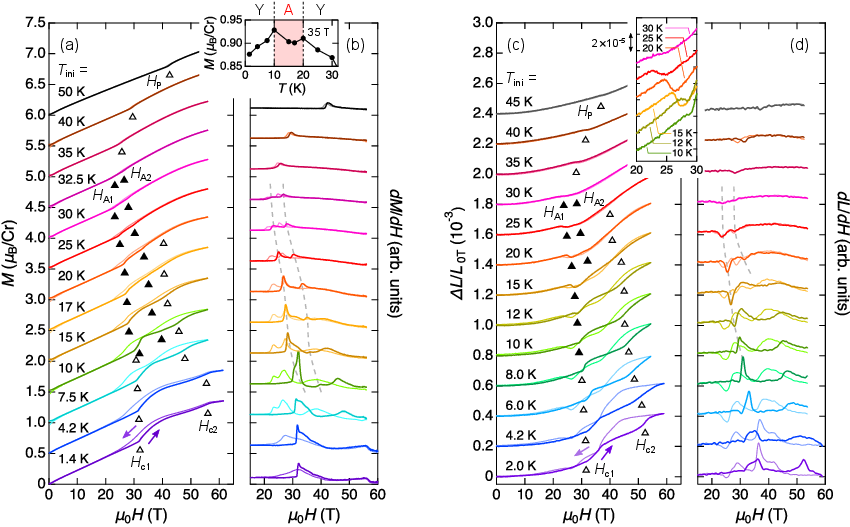}
\caption{(a) Magnetization, (b) its field-derivative, (c) longitudinal magnetostriction, and (d) its field-derivative as a function of magnetic field measured at various initial temperatures $T_{\rm ini}$. The thick (thin) lines correspond to the data in field-elevating (descending) processes. All the data except for the lowest-temperature one are shifted upward for clarity. The inset of (b) is a magnetization-vs-temperature plot at 35~T. The inset of (c) shows the enlarged view of the magnetostriction curves in field-elevating processes for 10~K $\leq T_{\rm ini} \leq$ 30~K, where lattice shrinkage occurs as the precursor of the phase transition from {\it X} to {\it A} phase at $H_{\rm A1}$. The filled (open) triangles in (a) and (c) denote the phase boundaries, $H_{\rm A1}$ and $H_{\rm A2}$ ($H_{\rm c1}$, $H_{\rm c2}$, and $H_{\rm P}$), which are determined from the $dM/dH$ and $dL/dH$ anomalies, respectively. The dashed gray lines in (b) and (d) trace the boundaries of {\it A} phase.}
\label{Fig3}
\end{figure*}

\subsection{\label{Sec3_2} Temperature dependence of magnetization, magnetostriction, and magnetocapacitance up to 60 T}

\begin{figure*}[t]
\centering
\includegraphics[width=0.95\linewidth]{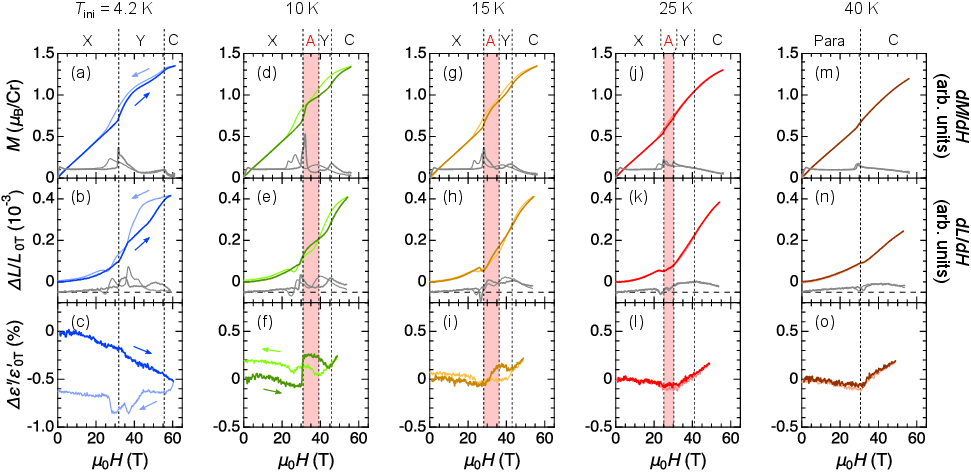}
\caption{Top, middle, and bottom panels show the magnetic-field dependence of magnetization, longitudinal magnetostriction, and magnetocapacitance, respectively, for (a)--(c) $T_{\rm ini}$ = 4.2~K, (d)--(f) 10~K, (g)--(i) 15~K, (j)--(l) 25~K, and (m)--(o) 40~K. The field-derivatives of magnetization and magnetostriction are also shown by gray lines in the right axes. The magnetocapacitance was measured along the field direction at a frequency of 50~kHz. The thick (thin) lines correspond to the data in field-elevating (descending) processes. The phase boundaries are roughly drawn by dashed black lines, which are determined from the data in the field-elevating process.}
\label{Fig4}
\end{figure*}

In order to clarify the in-field properties of CuInCr$_{4}$S$_{8}$, we measured magnetization, longitudinal magnetostriction, and magnetocapacitance in pulsed high magnetic fields of up to 60~T at various initial temperatures ($T_{\rm ini}$).
All the magnetization and magnetostriction data are summarized in Fig.~\ref{Fig3}, and some of them are arranged along with the magnetocapacitance data for several selected $T_{\rm ini}$ in Fig.~\ref{Fig4}.
Note that the observed results are irreversible and hysteretic especially for low $T_{\rm ini}$ because of significant sample-temperature changes in the field-descending process, as discussed in Sec.~\ref{Sec3_4}.
We will hence mainly focus on the data in the field-elevating process, which are useful for constructing the $H$-$T$ phase diagram (for the discussions on the field-descending process, see Appendix~\ref{SecF}).

For the lowest $T_{\rm ini}$ of 1.4~K, two-step metamagnetic transitions take place at $\mu_{0}H_{\rm c1}=32$~T and $\mu_{0}H_{\rm c2}=56$~T [Fig.~\ref{Fig3}(a)], followed by a 1/2-magnetization plateau which continues up to $\mu_{0}H_{\rm c3} \approx 110$~T \cite{2020_Gen}.
Correspondingly, a substantial lattice expansion is observed at $H_{\rm c1}$ and $H_{\rm c2}$ [Fig.~\ref{Fig3}(c)], suggesting the importance of the SLC on these phase transitions.
As discussed in Ref.~\cite{2020_Gen}, the overall magnetization curve can be roughly reproduced by the minimal Hamiltonian incorporating the SLC, which effectively produces the biquadratic exchange term $-({\mathbf S}_{i} \cdot {\mathbf S}_{j})^2$ between NN spins \cite{2004_Pen, 2021_Aoy}.
According to this, the spin configurations within each small tetrahedron in the low-field {\it X} phase, the intermediate-field {\it Y} phase between $H_{\rm c1}$ and $H_{\rm c2}$, and {\it C} phase with the 1/2-magnetization plateau can be assigned to the canted 2:2, canted 2:1:1, and collinear 3-up-1-down state, respectively [Fig.~\ref{Fig6}(b)].
As $T_{\rm ini}$ increases, each critical field is gradually decreased.
The hump structures at $H_{\rm c2}$ in $dM/dH$ and $dL/dH$ become blunt for $T_{\rm ini} \geq 20$~K [Figs.~\ref{Fig3}(b) and \ref{Fig3}(d)], making the phase boundary indiscernible.
For $T_{\rm ini} \geq 35$~K (=$T_{\rm N}$), the transition from the paramagnetic to a field-induced ordered phase is observed.
The critical field $H_{\rm P}$ becomes higher on increasing $T_{\rm ini}$ up to 50~K, suggesting that {\it C} phase is stabilized by thermal fluctuations and may extend to higher temperatures.
A moderate magnetocapacitance change is also seen at the transition into {\it C} phase for $T_{\rm ini} \geq 10$~K [Figs.~\ref{Fig4}(f), \ref{Fig4}(i), \ref{Fig4}(l), and \ref{Fig4}(o)], which may be attributed to the local lattice distortion to stabilize the 3-up-1-down magnetic structure via the exchange-striction mechanism.

\subsection{\label{Sec3_3} Magnetostrictive and magnetocapacitive effects in {\it A} phase}

In addition to the previously reported {\it X}, {\it Y}, and {\it C} phases, another field-induced phase ({\it A} phase) is found in the series of high-field experiments.
Let us here focus on the magnetization and magnetostriction curves for an intermediate $T_{\rm ini}$ range between 10 and 35~K.
Taking a closer look at Fig.~\ref{Fig3}(a), one can notice that the lowest-field metamagnetic transition becomes exceptionally steep for $T_{\rm ini} = 10$~K, which is clearly shown as the peak sharpness in the $dM/dH$ curve [Fig.~\ref{Fig3}(b)].
This is reflected in the plot of magnetization-versus-temperature at 35~T showing a maximum at 10~K [inset of Fig.~\ref{Fig3}(b)].
Besides, a dip structure of $\Delta L/L_{\rm 0T}$ appears in a narrow field range for 10~K $\leq T_{\rm ini} \leq$ 30~K [inset of Fig.~\ref{Fig3}(c)].
These features suggest that the nature of the lowest-field metamagnetic transition changes across $T_{\rm ini}=10$~K.
Accordingly, the critical field above 10~K should be distinguished from $H_{\rm c1}$ and instead termed $H_{\rm A1}$.
Importantly, another tiny $dM/dH$ peak appears at around the midpoint between $H_{\rm A1}$ and $H_{\rm c2}$ (termed $H_{\rm A2}$) for $T_{\rm ini}=10$~K, developing toward higher $T_{\rm ini}$ up to 32.5~K [Fig.~\ref{Fig3}(b)].
An anomaly is also visible at $H_{\rm A2}$ in the $\Delta L/L_{\rm 0T}$ data for 20~K $\leq T_{\rm ini} \leq$ 30~K [Fig.~\ref{Fig3}(c)].
These observations indicate the existence of another phase intervened between {\it X} and {\it Y} phases.

According to the microscopic magnetoelastic theory \cite{2019_Ros}, the classical Heisenberg antiferromagnet on the regular pyrochlore lattice exhibits a volume expansion, equivalently a positive magnetostriction, proportional to $M^{2}$ during the successive field-induced phase transitions until saturation, when assuming a constant 16-sublattice magnetic unit cell.
This tendency is roughly seen in the $\Delta L/L_{\rm 0T}$ data of CuInCr$_{4}$S$_{8}$ below 10~K [Figs.~\ref{Fig3}(a) and \ref{Fig3}(c)], where $\Delta L/L_{0T}$ follows a parabolic field-dependence within {\it X} phase and smoothly increases even across the phase transitions at $H_{\rm c1}$ and $H_{\rm c2}$.
This suggests that the magnetic unit cells in {\it Y} and {\it C} phases remain unchanged as in {\it X} phase.
From this viewpoint, the observed dip structure of  $\Delta L/L_{\rm 0T}$ for 12~K $\leq T_{\rm ini} \leq$ 30~K is unusual.
The negative magnetostriction is not seen within {\it A} phase, but occurs as a precursor to the metamagnetic transition from {\it X} to {\it A} phase at $H_{\rm A1}$ [Figs.~\ref{Fig4}(g), \ref{Fig4}(h), \ref{Fig4}(j), and \ref{Fig4}(k)].
We infer that such a nonmonotonic  $\Delta L/L_{\rm 0T}$ behavior would arise from the modification of the magnetic unit cell, i.e., the reconstruction of the $q$-vector.

Interestingly, it is uncovered that {\it A} phase is associated with a substantial magnetodielectric coupling in addition to the anomalous magnetostrictive behavior.
No signature of a magnetocapacitance change is observed at $H_{\rm c1}$ within our experimental resolution for $T_{\rm ini}=4.2$~K [Fig.~\ref{Fig4}(c)], while the magnetocapacitance is suddenly enhanced by $\sim$0.3\% at $H_{\rm A1}$ for $T_{\rm ini}=10$~K [Fig.~\ref{Fig4}(f)], which is exactly the characteristic temperature of the appearance of {\it A} phase.
Such a magnetocapacitive response is visible also for $T_{\rm ini}=15$ and 25~K [Figs.~\ref{Fig4}(i) and \ref{Fig4}(l)], signaling the emergence of multiferroicity in {\it A} phase originating from the helical spin modulation \cite{2005_Kat, 2007_Ari}.

\subsection{\label{Sec3_4} Sample-temperature change in our experimental condition with millisecond pulsed-field duration}

\begin{figure}[t]
\centering
\includegraphics[width=\linewidth]{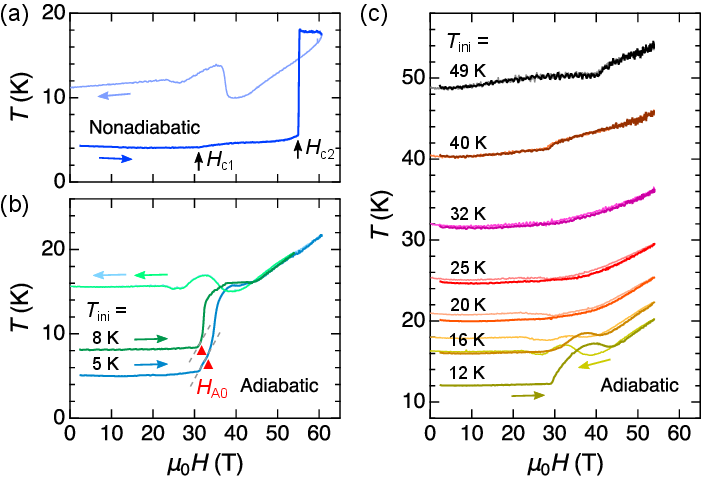}
\caption{$T(H)$ curves obtained (a) under the nonadiabatic condition at $T_{\rm ini}=4.2$~ K and (b, c) under the adiabatic condition at various $T_{\rm ini}$. The thick (thin) lines correspond to the data in the field-elevating (descending) process. In (b), the $T(H)$ curves in field-descending processes for $T_{\rm ini}=5$ and 8~K completely overlap with each other. Dashed gray lines are guides to the eye to make it easier to see the possible phase transition from {\it Y} to {\it A} phase at $H_{\rm A0}$ (denoted by red triangles).}
\label{Fig5}
\end{figure}

All the aforementioned measurements were performed under the nonadiabatic condition by immersing the sample in $^{4}$He liquid or gas.
However, the heating or cooling effects of the sample are usually nonnegligible in millisecond pulsed-field experiments \cite{2021_Miy}. 
Accordingly, we measured the sample temperature $T(H)$ in pulsed magnetic fields of up to 60~T under the nonadiabatic condition at $T_{\rm ini}= 4.2$~K.
Surprisingly, $T(H)$ was found to drastically increase by more than 10 K at $H_{\rm c2}$ in the field-elevating process [Fig.~\ref{Fig5}(a)].
This is possibly due to the hysteresis loss accompanied by the first-order phase transition into {\it C} phase.
Since the thermal equilibrium was not sufficiently achieved around the maximum field, the sample was not cooled down below 10~K in the field-descending process.
Here, successive phase transitions more complicated than those from {\it C}, {\it Y}, {\it A}, to {\it X} phase would take place, judging from at least three anomalies in the $T(H)$ curve in the field-descending process (Appendix~\ref{SecF}).
Note that such a strong heating effect does not seem to occur in the magnetization measurements at $T_{\rm ini}=1.4$ and 4.2~K because the magnetization curves in the field-descending process are less hysteretic and no double-hump structure is seen in the $dM/dH$ curves around 25~T unlike for $T_{\rm ini} \geq 7.5$~K [Figs.~\ref{Fig3}(a), (b), and \ref{Fig4}(a)].
The difference in the degree of the heating effect might be due to the sample form as well as the thermal bath environment; powder samples were used for the magnetization measurements, whereas sintered samples were used for other measurements.
We stress that a nearly isothermal condition is confirmed below $H_{\rm c2}$ for all the measurements at $T_{\rm ini}=4.2$~K.
This would be applicable also for $T_{\rm ini}>4.2$~K because the heating effect becomes less pronounced for higher $T_{\rm ini}$ as shown next.
Thus, it would be reasonable to use all the data in the field-elevating process for mapping out the $H$-$T$ phase diagram.

\subsection{\label{Sec3_5} Magnetic entropy changes through successive phase transitions}

In order to extract information about the magnetic entropy ($S_{\rm M}$) in each phase, we further investigated the magnetocaloric effect (MCE) under the adiabatic condition.
The MCE is the sample-temperature change induced by the evolution of $S_{\rm M}$ upon the application or removal of an external magnetic field.
When there is no irreversible first-order transition, the obtained $T(H)$ curve can be regarded as the isentropic curve.
As a consequence, an $S_{\rm M}$ increase (decrease) is detected as a $T(H)$ decrease (increase) \cite{2013_Kih}.

\begin{figure}[t]
\centering
\includegraphics[width=\linewidth]{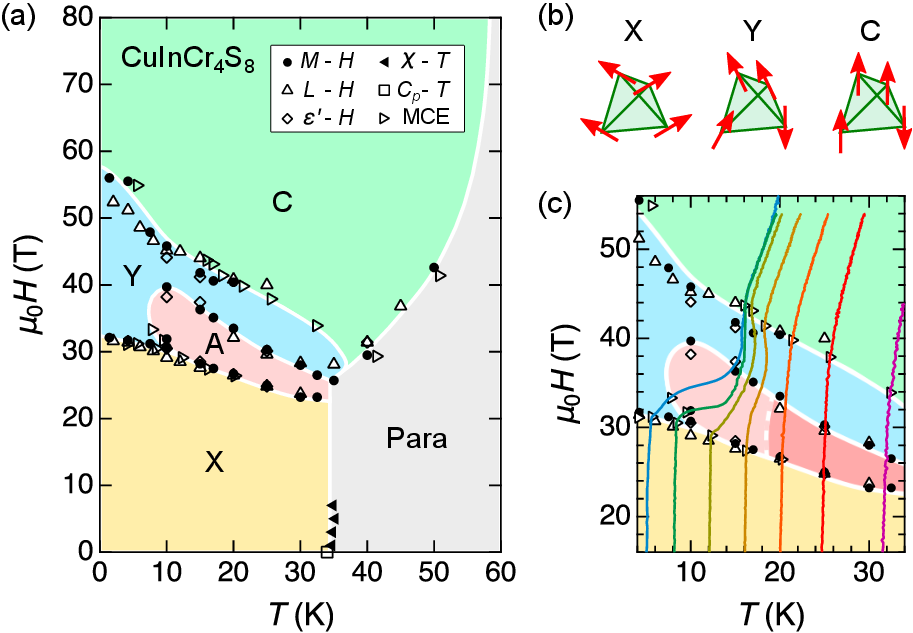}
\caption{(a) Magnetic-field-vs-temperature ($H$-$T$) phase diagram of CuInCr$_{4}$S$_{8}$ derived from a series of experiments in the present work. (b) Schematic of the magnetic structures in {\it X}, {\it Y}, and {\it C} phases predicted by the previous magnetoelastic theory \cite{2020_Gen}. (c) Adiabatic $T(H)$ curves in the field-elevating process shown in Figs.~\ref{Fig5}(b) and \ref{Fig5}(c) are plotted on the phase diagram, focusing on the $H$-$T$ region where multiple phase transitions take place. Judging from the magnetoentropic signatures, {\it A} phase is possibly further subdivided into two distinct phases in the low- (pink) and high-temperature regions (red).}
\label{Fig6}
\end{figure}

For $T_{\rm ini} \leq T_{\rm N}$, no apparent $T(H)$ change is observed below $H_{\rm c1}$ or $H_{\rm A1}$ [Figs.~\ref{Fig5}(b) and \ref{Fig5}(c)], indicating that there is almost no $S_{\rm M}$ change in {\it X} phase upon the application of a magnetic field.
Considering the magnetocaloric relation fulfilled under the isentropic condition
\begin{equation}
\label{MCE}
\left(\frac{\partial T}{\partial H}\right)_{S}=-\frac{T}{C_{H}}\left(\frac{\partial M}{\partial T}\right)_{H},
\end{equation}
where $C_{H}$ is the specific heat at the constant magnetic field, this observation is compatible with the $M/H$-$T$ data at a low field showing a slight temperature dependence below $T_{\rm N}$ [Fig.~\ref{Fig2}(a)].
For $T_{\rm ini}=5$ and 8~K, however, the $T(H)$ curves show a dramatic increase above $H_{\rm c1}$ and eventually overlap with each other around the maximum field [Fig.~\ref{Fig5}(b)].
Those in field-descending processes are completely irreversible, indicating that the contribution of the hysteresis loss is larger than that of an intrinsic MCE in this temperature range.
Importantly, an additional $T(H)$ kink is observed at $H_{\rm A0}$ just above $H_{\rm c1}$, presumably indicating the phase transition from {\it Y} to {\it A} phase [see dashed gray lines in Fig.~\ref{Fig5}(b)].
By drawing the $T(H)$ curves in field-elevating processes on the $H$-$T$ phase diagram, it can be seen that $T(H)$ significantly increases within {\it A} phase [Fig.~\ref{Fig6}(c)].
Bearing $(\partial M/\partial T)_{H}<0$ just above 10~K at around 35~T [inset of Fig.~\ref{Fig3}(b)] and Eq.~(\ref{MCE}) in mind, this would reflect the intrinsic entropy reduction in the low-temperature region of {\it A} phase.

As $T_{\rm ini}$ increases, the $T(H)$ change at $H_{\rm A1}$ and the accompanying hysteresis become smaller [Fig.~\ref{Fig5}(c)], assuring the achievement of the quasi-isentropic condition.
Notably, the $T(H)$ behaviors for 20~K $\leq T_{\rm ini} \leq$ 32~K are qualitatively different from those for $T_{\rm ini}= 12$ and 16~K; the inflection point at $H_{\rm A1}$ disappears for the former, indicating that $S_{\rm M}$ in {\it A} phase is as high as that in {\it X} phase above 20~K.
As discussed in Sec.~\ref{Sec3_1} and Appendix~\ref{SecE}, $S_{\rm M}$ in {\it X} phase still remains large just below $T_{\rm N}$ and gradually decreases toward low temperatures.
Considering this characteristic entropy nature of {\it X} phase together with the way of entropy changes across the transition from X- to {\it A} phase revealed by the MCE measurements, the difference in $S_{\rm M}$ between the low- and high-temperature sides of {\it A} phase should be significant.
Thus, we infer that {\it A} phase would be further subdivided into two phases due to the entropy effect, as shown in Fig.~\ref{Fig6}(c).
This may be related to the observation of the $\Delta L/L_{\rm 0T}$ anomaly at $H_{\rm A2}$ only for 20~K $\leq T_{\rm ini} \leq$ 30~K [Figs.~\ref{Fig3}(c) and \ref{Fig3}(d)].
At the transitions from {\it Y} and the paramagnetic phases to {\it C} phase, $T(H)$ commonly starts rising for
all the $T_{\rm ini}$ because of the lowest-$S_{\rm M}$ nature in the 3-up-1-down collinear magnetic structure [Figs.~\ref{Fig5}(b) and \ref{Fig5}(c)].
These observations are helpful for determining the phase boundary, especially between {\it Y} and {\it C} phases.

\subsection{\label{Sec3_6} ${\textit H}$--$\textit T$ phase diagram}

Based on the above experimental results, we construct the $H$-$T$ phase diagram of CuInCr$_{4}$S$_{8}$ for Sample no.~1 ($T_{\rm N}=35$~K), as shown in Fig.~\ref{Fig6}(a).
{\it A} phase is found to appear in a closed $H$-$T$ regime around 25-40~T and 10-35~K.
It should be noted that a similar phase diagram was identified for Sample no.~2 ($T_{\rm N}=32$~K), but {\it A} phase seems to further extend down to below 1.4 K (Appendix~\ref{SecC}).
These facts indicate that the appearance of {\it A} phase is an intrinsic property of CuInCr$_{4}$S$_{8}$, and its metastability is quite sensitive against the sample quality.

Although the perfect theoretical description which can reproduce the $H$-$T$ phase diagram of CuInCr$_{4}$S$_{8}$ is elusive at this stage, we propose one possible origin for the appearance of {\it A} phase.
In all the observed field-induced phase transitions, local lattice distortions as well as the lattice constant change should take place.
This induces a modulation of the strengths of $J$ and $J'$ from bond to bond; e.g., the trigonal distortion produces three $J(1+\delta_{1})$ and three $J(1-\delta_{2})$ per one small tetrahedron in the 3-up-1-down collinear state \cite{2011_Kim}.
It has been theoretically shown that the introduction of tetragonal or trigonal distortions into the FCC-lattice Heisenberg antiferromagnet can induce incommensurate helical orders at zero field \cite{1972_Yam}.
This may be related to the observed in-field properties of CuInCr$_{4}$S$_{8}$; the lattice distortion would lead to the competition between commensurate {\it Y} and incommensurate {\it A} phases, and the latter is stabilized on the higher temperature side by the entropy effect.

Finally, we note that the present assignment of {\it A} phase is tentative and it may be further subdivided into the low-$S_{\rm M}$ and high-$S_{\rm M}$ phases [Fig.~\ref{Fig6}(c)], as mentioned in Sec.~\ref{Sec3_5}.
If this is the case, the emergence of a multiple-$q$ state with a superposition of multiple spin modulations, such as a skyrmion lattice (SkL), is possible in the high-temperature side of {\it A} phase.
The SkL phase has been so far found in a number of materials with a noncentrosymmetric crystal lattice, i.e., MnSi and Cu$_{2}$OSeO$_{3}$ \cite{2009_Muh, 2012_Sek, 2021_Tok}.
There, the spin helix is induced by the competition between the FM exchange and DM interactions, and the triple-$q$ state is stabilized relative to the corresponding single-$q$ state by thermal fluctuations.
Indeed, the high-$S_{\rm M}$ nature of the SkL phase has been experimentally demonstrated by analyzing the magnetization data based on the magnetocaloric relation (\ref{Fig1}) \cite{2018_Boc}.
Accordingly, the SkL phase usually appears in a closed $H$-$T$ regime (often called an ``{\it A}-phase pocket'').
There also exist theoretical predictions on the realization of thermal fluctuation-driven multiple-$q$ states caused by the magnetic frustration, such as SkL (triple-$q$) in the $J_{1}$--$J_{3}$ model on the triangular lattice \cite{2012_Oku} and the meron/antimeron-like lattice (double-$q$) in the $J_{1}$--$J_{2}$ model on the honeycomb lattice \cite{2019_Shi_PRL, 2019_Shi_PRB}.
The proposed $H$-$T$ phase diagrams are rather complicated, similar to those of the DM-induced skyrmion hosts accommodating the {\it A}-phase pocket.
Although several compounds such as NiGa$_{2}$S$_{4}$ \cite{2005_Nak} and Bi$_{3}$Mn$_{4}$O$_{12}$(NO$_{3}$) \cite{2009_Smi} have been raised as candidates for realizing these predictions, the requirement of high-field experiments has challenged their verifications.
Only recently, MnSc$_{2}$S$_{4}$, characterized by the $J_{1}$--$J_{2}$--$J_{3}$ model on the centrosymmetric diamond lattice, was found to exhibit a fractional antiferromagnetic SkL, which seems to be attributed to the anisotropic exchange couplings as well as the magnetic frustration and is stabilized even at zero temperature \cite{2020_Gao}.
Furthermore, it has been theoretically proposed that the $J$--$J'$--$J_{3}$ model on the breathing pyrochlore lattice can host a quadrupole-$q$ hedgehog-lattice spin texture at zero field \cite{2021_Aoy_HG, 2022_Aoy_HG}.
In light of these backgrounds, it would be an intriguing subject whether or not the multiple-$q$ state is realized within {\it A} phase of CuInCr$_{4}$S$_{8}$.
It is necessary to perform NMR or neutron scattering experiments on single crystals, which are left for future work. Further theoretical investigations are also desired to verify the proposed scenario.

\section{\label{Sec4}Summary}

In summary, we have revealed that the breathing pyrochlore magnet CuInCr$_{4}$S$_{8}$, whose magnetism is governed by the magnetic frustration and the SLC, exhibits an exotic $H$-$T$ phase diagram with an {\it A}-phase pocket.
The appearance of {\it A} phase is found to be associated with anomalous magnetostrictive and magnetocapacitive responses.
This suggests the manifestation of helimagnetism in contrast to the neighboring commensurate magnetic phases.
Besides, the high-entropy nature is confirmed in the high-temperature side of {\it A} phase.
These features are reminiscent of the emergence of a thermal-fluctuations-driven multiple-$q$ phase such as the SkL.
To the best of our knowledge, there have been no reports on the $H$-$T$ phase diagram dressed with the {\it A}-phase pocket in insulating frustrated magnets.
We believe that the present discovery will attract more interest in breathing pyrochlore systems as well as open up new avenues for exploring multiple-$q$ states.

\section*{Acknowledgements}

The authors are grateful for helpful discussions with S. Gao, Y. Ishii, and S. Kitou.
The authors thank S. Kitani for providing the data of the specific heat of CdCr$_{2}$O$_{4}$ shown in Ref.~\cite{2013_Kit}.
This work was financially supported by the Japan Society for the Promotion of Science (JSPS) KAKENHI Grants-In-Aid for Scientific Research (No. 19H05823, No. 20J10988, No. 20K20892, and No. 22H04467), UTEC-UTokyo FSI Research Grant Program, and Basic Science Program No. 18-001 of the Tokyo Electric Power Company (TEPCO) memorial foundation.
M.G. was a postdoctoral research fellow of the JSPS.

\appendix

\section{\label{SecA}Structural analysis on Samples no.~1 and no.~2}

\begin{table}[b]
\renewcommand{\arraystretch}{1.2}
\caption{Crystallographic parameters of CuInCr$_{4}$S$_{8}$ polycrystalline samples for (a) sample no.~1 and (b) sample no.~2 at room temperature. The space group is $F{\overline 4}3m$.}
\begin{tabular}{ccccccc} \hline\hline
\multicolumn{7}{l}{(a) Sample no.~1 ($T_{\rm N} = 35$~K), $T = 300$~K,}\\
\multicolumn{7}{l}{~~~~~$R_{\rm wp}=4.829$, $R_{\rm p}=3.594$, $R_{\rm e}=3.325$, $S=1.4523$,}\\
\multicolumn{7}{l}{~~~~~$a=10.05867(3)$~\AA} \\ \hline
~ & ~ & $x$ & $y$ & $z$ & ~Occup.~ & B (\AA) \\ \hline
~~~Cu~~~ & ~~$4a$~~ & 0 & 0 & 0 & 1 & ~~0.935(47)~~~ \\
~~~In~~~ & ~~$4d$~~ & 3/4 & ~~3/4~~ & ~~3/4~~ & 1 & ~~0.425(19)~~~ \\
~~~Cr~~~ & ~~$16e$~~ & ~~0.36994(16)~~ & ~~$x$~~ & ~~$x$~~ & 1 & ~~0.364(26)~~~ \\
~~~S1~~~ & ~~$16e$~~ & ~~0.13341(23)~~ & ~~$x$~~ & ~~$x$~~ & 1 & ~~0.642(53)~~~ \\
~~~S2~~~ & ~~$16e$~~ & ~~0.61116(18)~~ & ~~$x$~~ & ~~$x$~~ & 1 & ~~0.432(37)~~~ \\ \hline\hline
\multicolumn{7}{l}{~~~}\\ \hline\hline
\multicolumn{7}{l}{(b) Sample no.~2 ($T_{\rm N} = 32$~K), $T = 300$~K,}\\
\multicolumn{7}{l}{~~~~~$R_{\rm wp}=3.945$, $R_{\rm p}=2.401$, $R_{\rm e}=1.608$, $S=2.4526$,}\\
\multicolumn{7}{l}{~~~~~$a=10.06270(3)$~\AA} \\ \hline
~ & ~ & $x$ & $y$ & $z$ & ~Occup.~ & B (\AA) \\ \hline
~~~Cu~~~ & ~~$4a$~~ & 0 & 0 & 0 & 1 & ~~1.321(44)~~~ \\
~~~In~~~ & ~~$4d$~~ & 3/4 & ~~3/4~~ & ~~3/4~~ & 1 & ~~0.636(17)~~~ \\
~~~Cr~~~ & ~~$16e$~~ & ~~0.37061(16)~~ & ~~$x$~~ & ~~$x$~~ & 1 & ~~0.621(24)~~~ \\
~~~S1~~~ & ~~$16e$~~ & ~~0.13334(20)~~ & ~~$x$~~ & ~~$x$~~ & 1 & ~~0.998(55)~~~ \\
~~~S2~~~ & ~~$16e$~~ & ~~0.61130(16)~~ & ~~$x$~~ & ~~$x$~~ & 1 & ~~0.554(33)~~~ \\ \hline\hline
\end{tabular}
\label{tab:Lietveld}
\end{table}

\begin{figure}[t]
\centering
\includegraphics[width=0.95\linewidth]{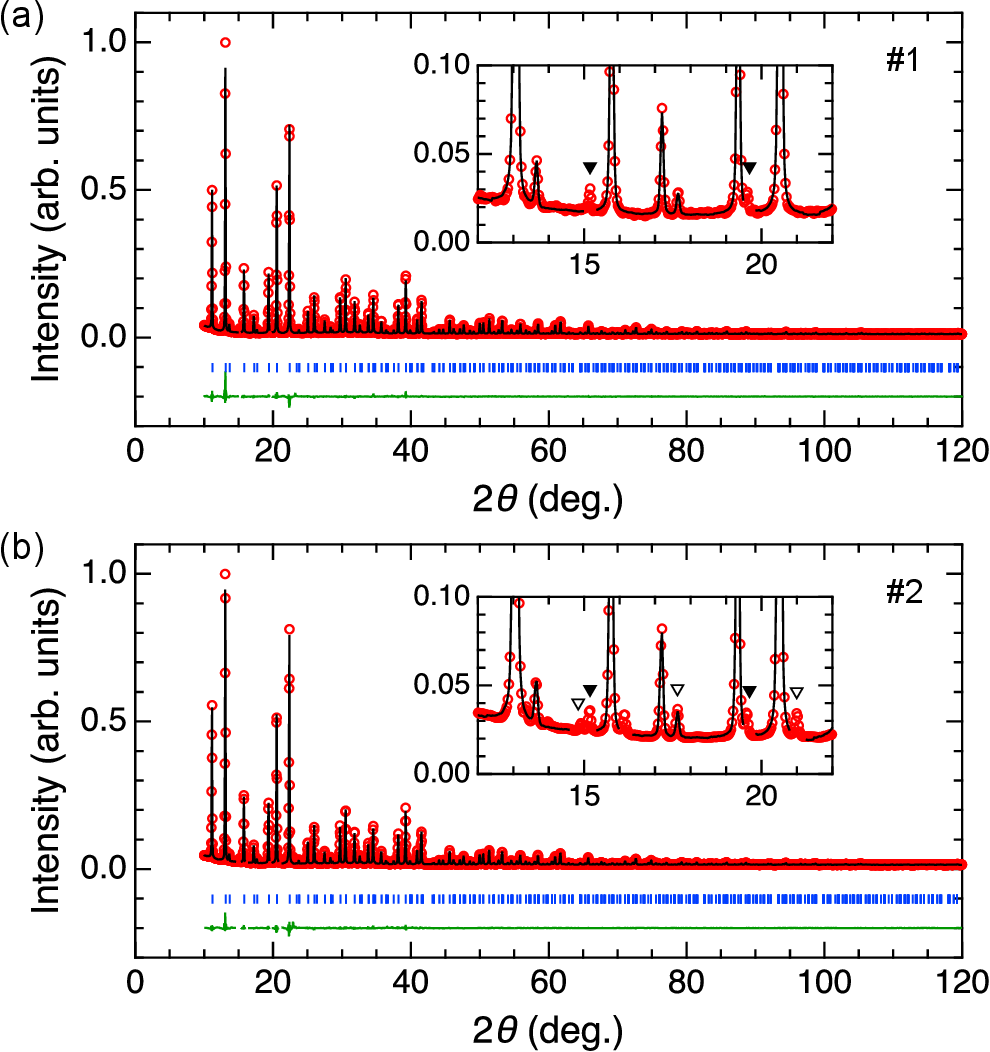}
\caption{Synchrotron powder XRD patterns of CuInCr$_{4}$S$_{8}$ taken for (a) Sample no.~1 and (b) Sample no.~2 at room temperature. Open red circles indicate the experimental data. The overplotted black curves indicate the calculated patterns with the cubic $F\overline{4}3m$ space group as the main phase. The vertical bars indicate the positions of the Bragg reflections. The bottom curves show the difference between the experimental and calculated intensities. The peaks in the insets indicated by filled and open triangles are reflection peaks of Cr$_{2}$S$_{3}$ and unknown impurities, respectively, which were excluded in the Rietveld analyses.}
\label{FigS1}
\end{figure}

In order to clarify the origin for the difference in the magnetic ordering temperature between Samples no.~1 ($T_{\rm N}=35$~K) and no.~2 ($T_{\rm N}=32$~K), we performed powder XRD measurements for these two samples at room temperature using synchrotron x-ray with $\lambda = 0.689739~\AA$ at BL-8A at Photon Factory. 
Figure~\ref{FigS1} shows the synchrotron powder XRD patterns and their Rietveld analyses performed using the RIETAN-FP program \cite{RIETAN}.
The crystallographic parameters obtained by assuming the occupancy of each site to be unity are summarized in Table~\ref{tab:Lietveld}.

Although no clear signatures of the site mixing between Cu and In or the S deficiency were confirmed for both samples, some differences in the crystallographic parameters were found between them: (i) the lattice constant $a$ of Sample no.~2 is larger by $\sim$0.04\% than that of Sample no.~1, (ii) the temperature factors $B$ in Sample no.~2 are larger than those in Sample no.~1, and (iii) the position $x$ of the Cr site in Sample no.~2 is closer to $x=3/8$ than that in Sample no.~1, indicating that the degree of breathing $r'/r$, which can be calculated by $r'/r=(0.5-x)/(x-0.25)$, is larger for Sample no.~1.
As shown in the insets of Fig.~\ref{FigS1}, both samples are found to contain a tiny amount of the Cr$_{2}$S$_{3}$ impurity phase, whereas the presence of unknown phases is seen only for Sample no.~2.
This should cause larger nonstoichiometry as well as crystallographic disorder in the main CuInCr$_{4}$S$_{8}$ phase for Sample no.~2, which would result in larger $a$, larger $B$, and smaller $r'/r$.
Thus, we conclude that the quality of Sample no.~1 is better than that of Sample no.~2.

\begin{figure*}[t]
\centering
\vspace{+0.5cm}
\includegraphics[width=0.88\linewidth]{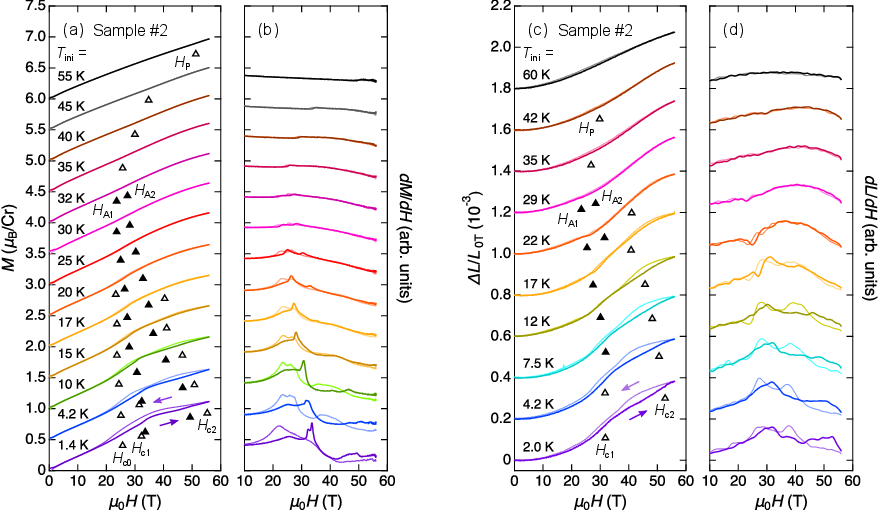}
\caption{(a) Magnetization, (b) its field-derivative, (c) longitudinal magnetostriction, and (d) its field-derivative of CuInCr$_{4}$S$_{8}$ as a function of magnetic field measured at various initial temperatures $T_{\rm ini}$ for Sample no.~2 ($T_{\rm N}$ = 32 K). The thick (thin) lines correspond to the data in the field-elevating (descending) process. All the data except for the lowest-temperature one are shifted upward for clarity. The filled (open) triangles in (a) and (c) denote the phase boundaries, $H_{\rm A1}$ and $H_{\rm A2}$ ($H_{\rm c0}$, $H_{\rm c1}$, $H_{\rm c2}$, and $H_{\rm P}$), which are determined from the $dM/dH$ and $dL/dH$ anomalies, respectively.}
\label{FigS2}
\end{figure*}

\section{\label{SecB}Magnetization and magnetostriction curves for Sample no.~2}

Figure~\ref{FigS2} summarizes the magnetization and magnetostriction data measured for Sample no.~2 at various initial temperatures ($T_{\rm ini}$) in pulsed magnetic fields of up to 57~T.
It can be seen that the observed phase transitions are much blunt compared to those for Sample no.~1 (Fig.~\ref{Fig3}) although all the critical fields $H_{\rm c1}$, $H_{\rm c2}$, $H_{\rm A1}$, $H_{\rm A2}$, and $H_{\rm P}$ are consistent with each other.
Furthermore, an additional $dM/dH$ hump is observed at around $\mu_{0}H_{\rm c0} \approx 25$~T for $1.4 \leq T_{\rm ini} \leq 20$~K [Figs.~\ref{FigS2}(a) and \ref{FigS2}(b)].
These observations signal the dispersion in the strength of exchange interactions caused by a tiny amount of crystallographic disorder in Sample no.~2 as mentioned above.

Note again that the magnetic long-range ordering temperature for Sample no.~2 ($T_{\rm N} = 32$~K) is lower than that for Sample no.~1 ($T_{\rm N} = 35$~K).
This would also originate from the crystallographic disorder in Sample no.~2.
We found that $T_{\rm N}$ changes slightly from synthesis to synthesis, and the magnetic properties are very sensitive to the sample quality.
As far as we have tried, Sample no.~1 adopted as the main data seems of the highest quality.

\section{\label{SecC}{\textit H}-{\textit T} phase diagram for Sample no.~2}

\begin{figure}[t]
\centering
\vspace{+0.5cm}
\includegraphics[width=0.65\linewidth]{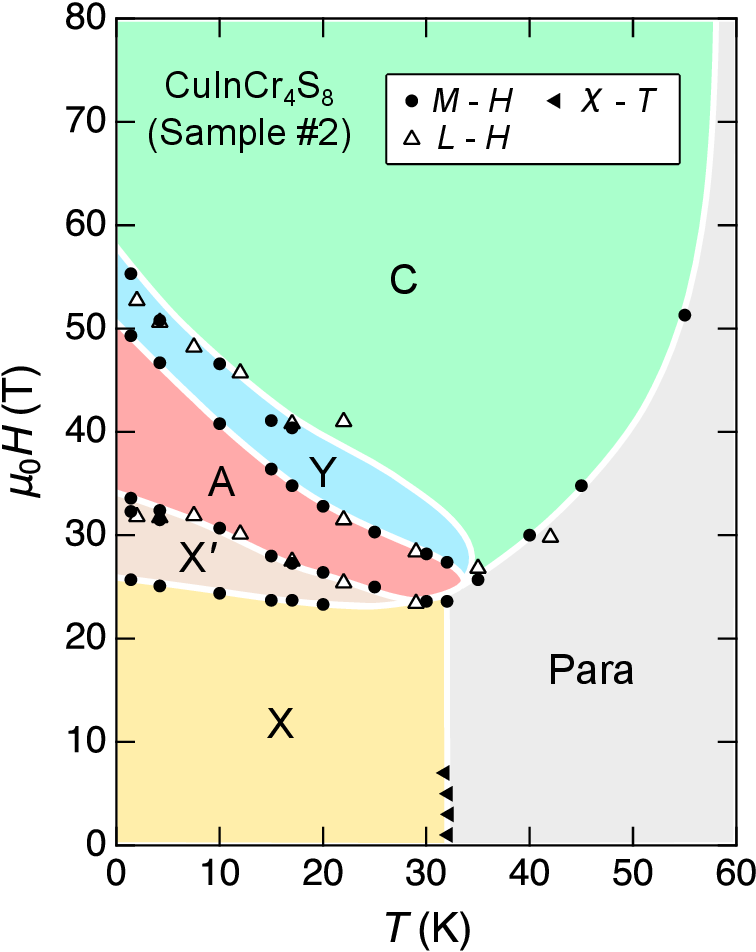}
\caption{$H$-$T$ phase diagram of CuInCr$_{4}$S$_{8}$ for Sample no.~2.}
\label{FigS3}
\end{figure}

Figure~\ref{FigS3} shows the $H$-$T$ phase diagram for Sample no.~2, which should have more crystallographic disorder than Sample no.~1 as discussed above.
The phase diagram is obtained from the magnetic susceptibility data up to 7~T and the high-field magnetization and magnetostriction data in the field-elevating process (Fig.~\ref{FigS2}).
The overall feature is similar with the $H$-$T$ phase diagram for Sample no.~1 [Fig.~\ref{Fig6}(a)], except for {\it A} and {\it X'} phases.
Importantly, {\it A} phase seems to further extend down to below 1.4 K, indicating that the appearance of {\it A} phase is an intrinsic property and its metastability is quite sensitive against the sample quality.

The existence of {\it X'} phase would not be an intrinsic property of CuInCr$_{4}$S$_{8}$ with the perfect breathing pyrochlore lattice and can be explained based on the following scenario.
The crystallographic disorder or nonstoichiometry locally distorts the breathing pyrochlore network, where Cr ions are aligned in a straight line along the [110] or equivalent direction.
As a consequence, the AFM exchange interaction between the NN Cr ions would be partially weakened because the overlap integral between the $t_{\rm 2g}$ orbitals decreases.
Therefore, the critical field of the first metamagnetic transition from {\it X} phase to {\it Y} or {\it A} phase would decrease in some parts of the crystal.
This results in the hump anomaly in $dM/dH$ at $H_{\rm c0}$ observed only in Sample no.~2 [Fig.~\ref{FigS2}(b)].
Accordingly, we attributed the emergence of {\it X'} phase to a phase coexistence of {\it X} and {\it A} (or {\it Y}) phases rather than a transition into a distinct phase.

\section{\label{SecD}DM interactions in the breathing pyrochlore lattice}

\begin{figure}[b]
\centering
\vspace{+0.5cm}
\includegraphics[width=0.95\linewidth]{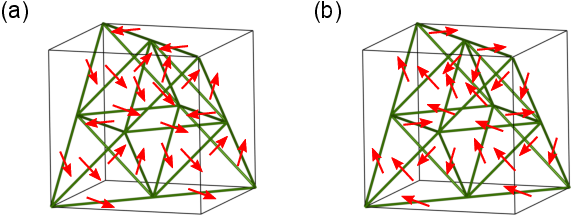}
\caption{Two possible DM vectors on the effective FCC lattice.}
\label{FigS4}
\end{figure}

For the regular pyrochlore lattice, Dzyaloshinskii--Moriya (DM) interactions can be present on all the NN bonds because of the absence of an inversion center at the middle point between NN sites \cite{1957_Dzy, 1960_Mor}.
As pointed out in Ref.~\cite{2005_Elh}, there are two possible ways to locate DM vectors, both of which resolve the macroscopic degeneracy caused by geometrical frustration and lead to the magnetic long-range order: one selects the ``all-in-all-out'' spin configuration whereas the other selects the coplanar one as the ground state.

When one introduces the breathing bond alternation on the pyrochlore lattice, DM vectors with different strength and sign would be allowed in the large and small tetrahedra.
In the present case of AFM $J>0$ and strong FM exchange $J'<0$, however, the effects of DM vectors within the large tetrahedra should be negligible, so that there are eventually two possible ways to locate effective DM vectors, which are depicted on the effective FCC lattice in Fig.~\ref{FigS4}.
The degeneracy within the type-I magnetic order with ${\mathbf q}$=(1 0 0) cannot be resolved by DM interactions in either case.
Therefore, we conclude that DM interactions are not responsible for the appearance of {\it X} phase.

\section{\label{SecE}Magnetic entropy changes below {\textit T}$_{\mathbf N}$ at zero field}

\begin{figure}[b]
\centering
\vspace{+0.5cm}
\includegraphics[width=\linewidth]{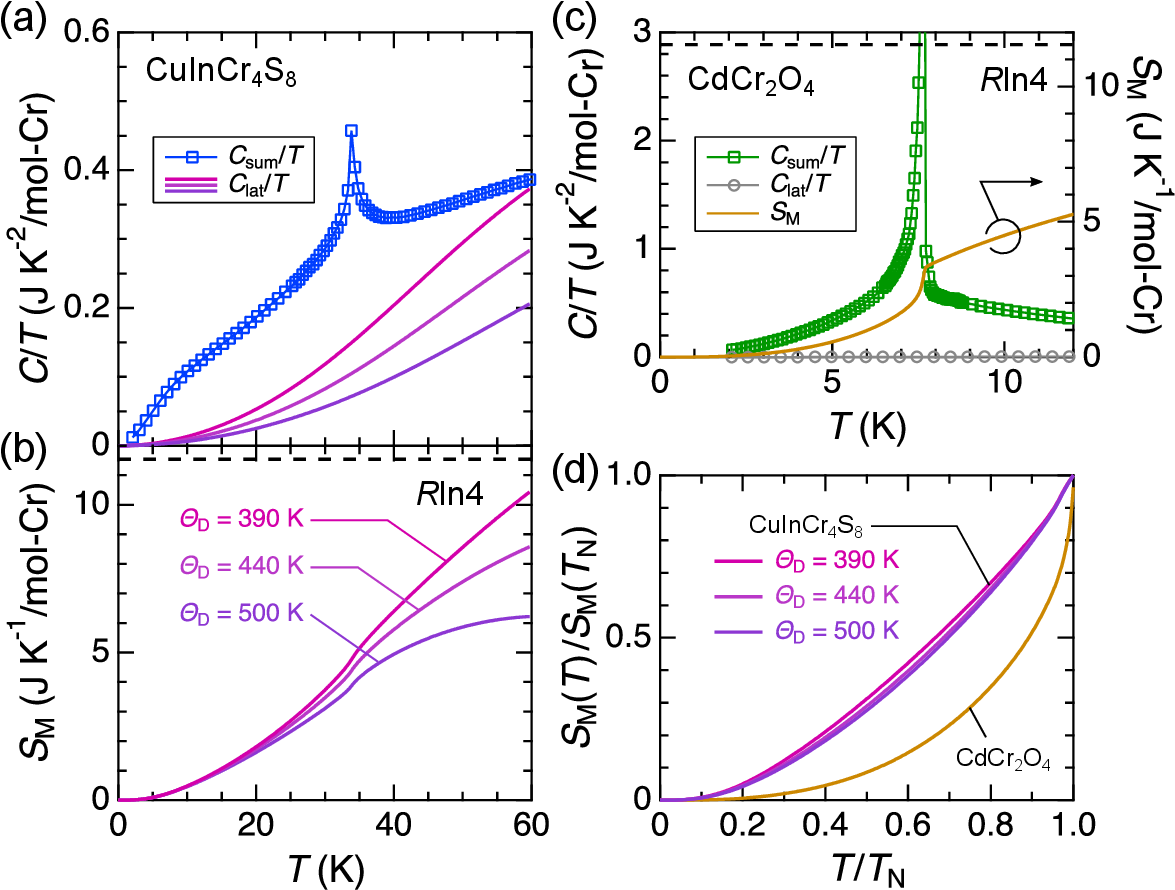}
\caption{(a) Temperature dependence of the specific heat divided by temperature $C_{\rm sum}/T$ for CuInCr$_{4}$S$_{8}$ at 0~T [Fig.~2(b)] and the estimated lattice contribution $C_{\rm lattice}/T$ based on the Debye model with $\Theta_{\rm D}=390, 440$, and 500~K. (b) Temperature dependence of the magnetic entropy for CuInCr$_{4}$S$_{8}$ calculated by integrating $(C_{\rm sum}-C_{\rm lattice})/T$. (c) Temperature dependence of the specific heat divided by temperature for CdCr$_{2}$O$_{4}$ and the nonmagnetic ZnGa$_{2}$O$_{4}$ at 0~T (left), and the calculated magnetic entropy for CdCr$_{2}$O$_{4}$ (right). The data are taken from Ref.~\cite{2013_Kit}. (d) Relative magnetic entropy changes below $T_{\rm N}$ for CuInCr$_{4}$S$_{8}$ and CdCr$_{2}$O$_{4}$.}
\label{FigS5}
\end{figure}

We here estimate the temperature dependence of the magnetic entropy in CuInCr$_{4}$S$_{8}$ at zero field based on the observed $C/T$--$T$ curve [Fig.~\ref{Fig2}(b)] and compare the results with that in the typical Cr spinel antiferromagnet CdCr$_{2}$O$_{4}$.
For estimating the lattice contribution, we adopt the Debye model due to the lack of a suitable nonmagnetic reference compound.
The calculated lattice contributions with the Debye temperature $\Theta_{\rm D}=390, 440$, and 500~K are shown in Fig.~\ref{FigS5}(a).
The magnetic entropy at $T_{\rm N}=35$~K is obtained as $S_{\rm M}=4 \sim 5$~JK$^{-1}$/mol-Cr [Fig.~\ref{FigS5}(b)], which is less than half of $R$ln4, indicating that the magnetic short-range order develops above $T_{\rm N}$.
In contrast to CuInCr$_{4}$S$_{8}$, the $C/T$--$T$ curve for CdCr$_{2}$O$_{4}$ exhibits a sharp peak at $T_{\rm N}=7.8$~K [Fig.~\ref{FigS5}(c)] \cite{2013_Kit}, where a structural transition also takes place \cite{2019_Ros}.
A similar sharp peak in $C/T$ associated with a strong entropy drop is the universal property in other Cr spinels \cite{2013_Kem} and the breathing pyrochlore families \cite{2018_Oka, 2013_Oka}.

Figure~\ref{FigS5}(d) visualizes the difference in the way of the magnetic entropy release below $T_{\rm N}$ between CuInCr$_{4}$S$_{8}$ and CdCr$_{2}$O$_{4}$. As readily seen, the entropy reduction just below $T_{\rm N}$ is much slower for CuInCr$_{4}$S$_{8}$ than for CdCr$_{2}$O$_{4}$: $S_{\rm M}(0.8T_{\rm N})/S_{\rm M}(T_{\rm N})\approx 0.64$ for CuInCr$_{4}$S$_{8}$, whereas $S_{\rm M}(0.8T_{\rm N})/S_{\rm M}(T_{\rm N})\approx0.35$ for CdCr$_{2}$O$_{4}$.
This supports the existence of spin fluctuation in {\it X} phase for CuInCr$_{4}$S$_{8}$, which would be originated from the magnetic frustration inherent in the effective FCC lattice.

\section{\label{SecF}Phase transitions during the field-descending process}

\begin{figure}[H]
\centering
\vspace{+0.5cm}
\includegraphics[width=0.75\linewidth]{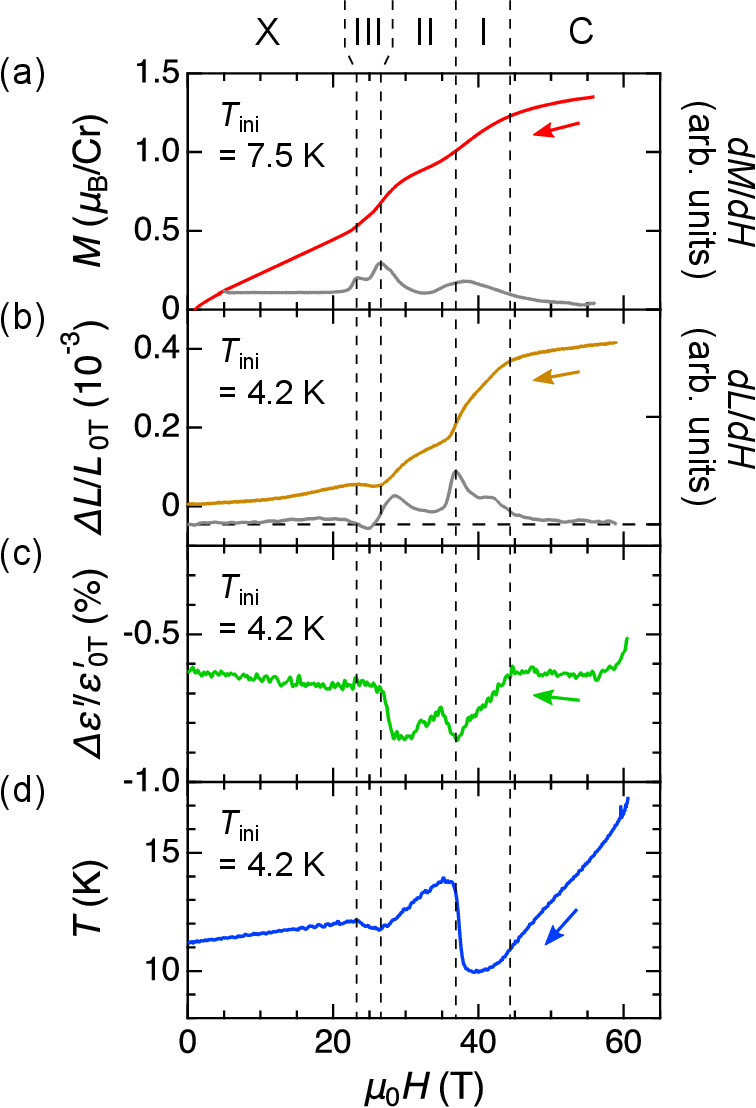}
\caption{(a) Magnetization, (b) longitudinal magnetostriction, (c) magnetocapacitance, and (d) magnetocaloric effect curves in the field-descending process measured for Sample no.~1. The field-derivatives of magnetization and magnetostriction are also shown by gray lines in the right axes.}
\label{FigS6}
\end{figure}

We here mention the nature of the successive phase transitions of CuInCr$_{4}$S$_{8}$ for the field- descending process.
Figure~\ref{FigS6} shows the magnetization, magnetostriction, magnetocapacitance, and magnetocaloric effect (MCE) curves in the field-descending process obtained under the nonadiabatic condition for Sample no.~1.
The latter three data are taken at $T_{\rm ini}=4.2$~K, while the magnetization are taken at $T_{\rm ini}=7.5$~K.
As discussed in the main text, CuInCr$_{4}$S$_{8}$ significantly suffers from the MCE by the pulsed-field application, so that the sample temperature during the field-descending process would be largely different from $T_{\rm ini}$ and dependent on the experimental setup.
Although the exact sample temperature is unknown in Figs.~\ref{FigS6}(a)--\ref{FigS6}(c), it would be roughly as shown in Fig.~\ref{FigS6}(d).
Remarkably, there exist at least four anomalies signaling phase transitions.
Similar to the case of the field-elevating process, the low-field phase below $\sim$23 T and the high-field phase above $\sim$44 T should correspond to {\it X} and {\it C} phases, respectively.
For the intermediate-field region, several kinds of magnetic phases beyond {\it A} and {\it Y} phases seem to appear accompanied by considerable magnetostrictive and magnetocapacitive responses as well as moderate entropy changes.

Here, we tentatively define them as phase I, II, and III from the high-field side. Judging from the dip structure of $\Delta L/L_{\rm 0T}$ and the sharp magnetocapacitive anomaly around the phase boundary between phases II and III, a large modification of the magnetic structure with the reconstruction of the $q$-vector would occur.

\end{document}